\documentclass[traditabstract]{aa}
\usepackage{graphicx,natbib}
\usepackage{epsfig}
\usepackage{aalongtable}
\usepackage{rotating}
\usepackage{amsmath,amssymb,color,latexsym,longtable,array}
\usepackage{txfonts}
\bibpunct{(}{)}{;}{a}{}{,} 
\begin{document}
\title{Polarimetry of optically selected BL Lac candidates from the SDSS
\thanks{Based on observations collected with 
the NTT on La Silla (Chile) operated by the European Southern
Observatory in the course of the observing proposal 082.B-0133}
\fnmsep
\thanks{Based on observations collected at the Centro Astron\'{o}mico
  Hispano Alem\'{a}n (CAHA), operated jointly by the
  Max-Planck-Institut f\"ur Astronomie and the Instituto de
  Astrofisica de Andalucia (CSIC)}
\fnmsep
\thanks{Based on observations made with the Nordic Optical Telescope,
operated on the island of La Palma jointly by Denmark, Finland,
Iceland, Norway, and Sweden, in the Spanish Observatorio del 
Roque de los Muchachos of the Instituto de Astrofisica de 
Canarias.}
\fnmsep
\thanks{Table 1 is only available in electronic form at the CDS via anonymous 
ftp to cdsarc.u-strasbg.fr (130.79.128.5) 
or via http://cdsweb.u-strasbg.fr/cgi-bin/qcat?J/A+A/}
}
\author{J. Heidt\inst{1}
          \and
	K. Nilsson\inst{2}}

\offprints{J. Heidt,\\
\email{jheidt@lsw.uni-heidelberg.de}}

\institute{ZAH, Landessternwarte Heidelberg, K\"onigstuhl, D-69117
     Heidelberg, Germany
\and
Finnish Centre for Astronomy with ESO (FINCA), University of Turku, 
V\"ais\"al\"antie 20, FI-21500 Piikki\"o, Finland}

\date{Received 19 January 2011 / Accepted 18 February 2011}


  \abstract
   {
We present and discuss polarimetric observations of 182
targets drawn from an optically selected sample of 240 
probable BL Lac candidates out of the
SDSS compiled by Collinge et al. (2005). In contrast
to most other BL Lac candidate samples extracted from the SDSS, its radio-
and/or X-ray properties have not been taken into account for its
derivation. Thus, because its selection is based on optical properties
alone, it may be less prone to selection effects inherent in
other samples derived at different frequencies, so it offers a
unique opportunity to extract the first unbiased BL Lac
luminosity function that is suitably large in size. 

We found 124 out of 182 targets (68\%) to be polarized, 
95 of the polarized targets (77\%) to be highly polarized ($>$ 4\%).
The low-frequency peaked BL Lac candidates in the sample are
on average only slightly more polarized than the high-frequency 
peaked ones. Compared to earlier studies, we
found a high duty cycle in high polarization ($ \sim
66^{+2}_{-14}\%$ to be  
$> 4\%$ polarized) in high-frequency peaked BL Lac candidates.  
This may come from our polarization analysis, which minimizes the
contamination by host galaxy light.

No evidence of radio-quiet BL Lac objects in the sample was found. 

Our observations show that the probable sample of BL Lac candidates of
Collinge et al. (2005) indeed contains a large number of bona fide BL
Lac objects. High S/N spectroscopy and deep X-ray observations are
required to construct the first luminosity function 
of optically selected BL Lac objects and
to test more stringently for any radio-quiet BL Lac
objects in the sample.}


\authorrunning{}
\titlerunning{
Polarimetry of optically selected SDSS BL Lac candidates
}
\maketitle

\section{Introduction}\label{introduction}

BL Lac objects are characterized by large variability from radio
to TeV frequencies, nearly featureless optical spectra, high
and variable polarization, and in many cases superluminal motion. They
reside in the nuclei of giant elliptical galaxies, where according to
the current paradigm, a supermassive black hole is accreting material
from its surroundings and collimating two relativistic jets in
opposite directions. According to the so-called ``Unified Scheme''
\citep{1995PASP..107..803U}, BL Lac objects are FR I radio galaxies with
one of the jets nearly pointing towards us. Relativistic effects can
boost the jet emission to a level where it almost completely
outshines the host galaxy.

Due to their extreme properties it is not surprising that BL Lacs form
$<$ 1\% of the entire AGN population known today. 
More than 100000 confirmed QSO's \citep{2010AJ....139.2360S} 
and more than $10^6$ QSO candidates \citep{2009ApJS..180...67R} have been published,
but the number of BL Lac (candidates) hardly exceeds 1000 
\citep{2010A&A...518A..10V}. 

BL Lac objects are classically
detected in radio- or X-ray surveys and have traditionally been divided
into two groups according to the location of the peak of their
synchrotron emission: the low-energy-peaked BL Lacs (LBL)
and high-energy-peaked BL Lacs (HBL). Until now, the resulting 
samples based on single surveys alone only contain a few dozen objects, 
irrespective of whether they have been compiled from radio 
\citep[e.g. the 1Jy sample by][]{1991ApJ...374..431S} or from 
X-ray observations 
\citep[e.g. the Einstein Slew Survey sample by][]{1996ApJS..104..251P}.
Samples compiled from a combination of both,
e.g., the ROSAT All-Sky Survey-Green Bank sample by
\citet[][RGB]
{1999ApJ...525..127L} and the Sedentary Survey by
\citet[][]{2005A&A...434..385G} contain a much larger  number of objects 
(100 - 200). Most important is that they contain
BL Lac objects midway between LBL and HBL.
Regardless of the selection criteria, all samples suffer from a 
relatively high number of up to 50\% of
sources with unknown or highly uncertain redshift
and are subject to various biases by their selection criteria.
BL Lacs detected in radio surveys are typically more core-dominated
than those detected in X-rays, and the surveys in different 
wavelength regimes have different depths.

Claims have been made that 
HBL and LBL may evolve differently
\citep{1991ApJ...374..431S,1991ApJ...380...49M}, but these claims are
subject to low number statistics and to the biases mentioned above,
which cannot be easily corrected for. 
Likewise, the general trend in the cosmic evolution of BL
Lacs is not constrained, with claims of negative, positive, or no
evolution \citep[see e.g.][]{2003A&A...401..927B,2007ApJ...662..182P}.
As a consequence, attempts to construct, say, a
luminosity function for BL Lacs and/or their hosts
were limited by small numbers \citep{2007ApJ...662..182P}.
The former is especially important since it allows a
test of the Unified Scheme and can constrain models for jet opening angles as
a function of luminosity by comparing it with the luminosity functions
of other radio-loud AGN.

A viable alternative would be to construct  a suitable 
optically selected sample of BL Lac objects. Since the optical
region lies between the HBL and LBL peak frequencies,
optically selected samples are
representative of the whole BL Lac population and may be subject 
to biases that are easier to control.
However, besides the obvious advantage of optically selected BL Lac samples
over the ones selected from radio and/or X-ray surveys, 
only a few attempts have been made to extract an optically
selected sample of BL Lacs.  Exactly because BL Lac objects share
optical properties with other sources 
\citep[e.g. featureless spectra,
variability, and linear polarization as in the case of magnetic DC white dwarfs,][] 
{1978ARA&A..16..487A}, it is very difficult to select and to
confirm candidates from optical data alone. 

Early attempts to detect BL Lac objects via optical properties
\citep[e.g.][]{1982MNRAS.201..849I,1984ApJ...276..449B,1993ApJ...404..100J} 
have only been moderately successful. Even with the more sophisticated
approach by \citet{2002MNRAS.334..941L}, who
extracted a sample of 56 featureless blue continuum sources with absent
proper motion from the 2-degree field QSO redshift survey
\citep[2QZ,][]{2004MNRAS.349.1397C}, only few a BL Lac objects have been
detected. Follow-up spectroscopy and NIR-imaging has revealed that 
most of the sources are
either stellar or extragalactic with faint, but broad emission
features. Only a very few good BL Lac candidates remain
\citep{2005A&A...434..895N,2007MNRAS.374..556L}. 
On the other hand,
\citet{2004MNRAS.352..903L} have found an intriguing object within their
sample, which could potentially be a radio-quiet BL Lac object, a class
of objects not believed to exist
\citep[e.g.][]{1990ApJ...348..141S}. This demonstrates the new
discovery space among optically selected samples.

The Sloan Digital Sky Survey (SDSS) offers a unique database 
for constructing 
such a sample. It is a multi-institutional effort to image 
10000 deg$^2$ on the
sky of the north galactic cap in 5 optical filters covering 3800 -
10000 \AA with follow-up moderate-resolution ($\lambda/\Delta \lambda
\sim 1800$) multi-object spectroscopy of about $10^6$ galaxies,
$10^5$ quasars, and a similar number of unusual objects 
\citep[see, e.g.][]{2000AJ....120.1579Y}. The SDSS uses a dedicated
2.5m telescope at Apache Point Observatory with a mosaic of 30 CCD cameras
providing a 2\fdg5 field of view \citep[][]{1998AJ....116.3040G}
as well as two multi-object fiber-fed spectrographs allowing to 640 spectra
to be taken simultaneously across a 7\degr field of view 
\citep[see, e.g.][]{2002AJ....123..485S}.

Compilations have already been presented by
\citet{2003AJ....126.2209A}, \citet[][A07
  hereafter]{2007AJ....133..313A}, and 
\citet{2008AJ....135.2453P} but here a cross-correlation with 
radio- and/or X-ray properties have been
considered for their derivation.
\citet[][C05 hereafter]{2005AJ....129.2542C} were the first to
present an optically selected sample of 386 BL Lac candidates out of
the SDSS, where
radio- or X-ray properties were a priori not taken into 
account\footnote{ ``A priori'' refers to C05 not
  cross-correlating his targets with X-ray or radio data bases {\em
    before} he extracted the catalog. In fact, 55 of the C05 BL Lac
  candidates were
  spectroscopically targeted by the SDSS due to their radio/X-ray
  properties. The remaining ones are mainly included in the SDSS
  spectroscopic database because of their UV excess 
\citep[see discussion in
  C05, section 5.5 and][section 2]{2007ApJ...663..118S},
so the C05 sample is not completely free of biases.}.
The compilation is divided into a set
of 240 probable and 146 possible candidates each. 
Recently, \citet[][P10a hereafter]{2010AJ....139..390P} have presented an
optically selected compilation of 723 BL Lac candidates from the SDSS data
release 7. His approach was similar to the one by C05 (and in
fact, P10a ``recovered'' 226 of the 240 probable BL Lac candidates from 
C05).

In order to extract the ``bona fide'' BL Lac objects in the 
probable sample of C05, we carried  out an extensive program to search for 
the two main characteristic properties of BL Lac objects among 
the sample, namely variability and polarization. 
In a first study of a subset of the sample, 
\citet[][S07 hereafter]{2007ApJ...663..118S}
found 24 out of  42 sources to be polarized.
In this paper, we enlarge the study by S07,  
present our data set and describe the polarization properties of 
182 out of 240 targets of the probable sample of BL Lac candidates of C05.
In combination with the variability characteristics and  host galaxy 
properties derived from our data, and
broad-band optical-NIR SEDs \citep[using data from the United Kingdom
Infrared Telescope Infrared Deep Sky Survey UKIDSS,][]{2007MNRAS.379.1599L} 
all of which 
will be determined in a forthcoming paper (Nilsson et al. in prep.), 
we will be moving towards the first well-defined optically selected sample 
of BL Lac objects unbiased with respect to its radio- or X-ray properties.

In the following sections, we present our observations and describe 
the data reduction, followed by analysis and a summary of the results.
We finally end with a discussion, some conclusions and further prospects.
Throughout this paper we use SDSS-magnitudes (which are very
  close to AB magnitudes) and a standard cosmology
with $H_0=70$ km/s/Mpc, $\Omega_M=0.3$, and $\Omega_{\Lambda}=0.7$.


\section{Sample selection, observations, and data reduction}

C05 extracted
  his sources from a set of over 345,000 individual SDSS spectra
  covering 2860 deg$^2$ on the sky \citep[roughly the area covered by the
  SDSS Data Release Two, see][]{2004AJ....128..502A}.

For the derivation of the sample, C05 selected quasi-featureless
spectra with the requirement of an S/N of at least $>$ 100 in one
of three spectral regions centered at 4750, 6250, and 7750 \AA. To get
rid of as many likely stellar contaminants as possible (mainly 
weak-featured white dwarfs), candidates with significant proper motion
were removed. This resulted in a catalog of 386 BL Lacertae
candidates, which were separated into two subsets with 240 
probable and 146 possible
candidates based on their optical colors and other properties, respectively.
``Probable'' signifies a probable 
extragalactic nature ($g - r \geq 0.35$ or $r - i \geq 0.13$, or X-ray
or radio counterpart, or measured redshift), while ``possible'' 
signifies a likely stellar nature ($g - r \leq 0.35$ and $r - i \leq 0.13$
and no indication of extragalactic nature). There are therefore good
reasons to believe that the contamination by stars in the ``probable''
list is very low, while stars are expected to dominate in the 
``possible'' list. 

For our project, we selected the catalog containing the 240 probable 
BL Lacs of C05 since a) 
most objects for which radio- and X-ray information is available 
cover the region in $\alpha_{\rm ro}$ - $\alpha_{\rm ox}$ space typical of
BL Lacs; b) it contains a 
large enough number of candidates for deriving a clean, optically 
selected sample, on one hand, but manageable in terms of telescope
time for our polarimetric observations, on the other; c)
it contains a suitable number of radio-weak BL Lac candidates; and d)
only 84 out of the 240 targets in the probable 
catalog were listed as BL Lac in NED before C05 published their 
sample{\footnote{In the following, ``listed in NED'' refers to
  ``listed as BL Lac in NED'' when C05 published their sample.}.
Redshifts are available for $>$ 50\% of the
sources (with the majority of them between z = 0 and 1.2).}

The observations were carried out during 16 nights spread over 4 runs
with the ESO-NTT (NTT), the Calar Alto 2.2m (CA), and
Nordic Optical Telescope (NOT) telescopes. 
The goal was to observe all 240 sources of the C05 sample, but
since not all nights could be used, we had to prioritize our observations.
Since the observations were scheduled on fixed dates, the selection of
the targets was primarily driven by RA constraints.
Whenever possible, highest priority was given to sources not listed
in NED at a given RA-range before moving to targets without
polarization measurements in the literature. A couple of sources were
observed with two different telescopes to check the reliability of our
analysis.

The observations at the NTT were split into
two separate runs, four nights each, from Oct 2 - 6, 2008 and from Mar 28
- April 1,
2009, respectively. Data could be acquired during 2 1/2 photometric
nights in October and during all four nights (3 of them photometric) 
in March/April. Seeing was mostly good (0\farcs6-1\farcs2) during 
both runs. We used EFOSC2 attached
to the NTT. The observations were taken through a Gunn-r filter 
(\#786), which matched the r'-filter used for the SDSS closest. For the
polarimetric observations, we used the Wollaston prism with a beam
separation of 10\arcsec\ and a 
half-wave plate. A 2k Loral-CCD with binning = 2 was employed,
which gave us a field of view of 4\arcmin $\times$ 4\arcmin\
(0\farcs24/pixel). This allowed a suitable number of stars in the field 
to be observed at
the same time for characterizing interstellar polarization
along the line of sight. As in all runs, we took care to place the
target at the center of the field of view. 
For each target, we took from one to four
sequences at PA = 0, 22.5, 45, and 67.5\degr. Exposure times 
ranged from 10 - 1000 sec per PA depending on the brightness of
the source. As in all runs, the goal was to obtain an S/N of $\geq$ 100
to reach an accuracy of 1\% or better. Since EFOSC2 was mounted at the
Nasmyth focus of the NTT, instrumental polarization varying as a
function of the position of the telescope on the sky was
expected. Thus,  five to six times per night an unpolarized
standard star from \citet{2007ASPC..364..503F} and three to four times per
night a polarized standard star provided at the ESO-WEB was observed. The
observations of the standard stars were spread homogeneously across the night.

The observations on CA were carried out in service 
mode using CAFOS attached to the 2.2m telescope during 
five photometric nights with good seeing ($\leq$ 1\farcs5) on
Feb 18 - 24, 2009. Again, a Wollaston prism with a beam separation of 
19\arcsec\  and a rotatable $\lambda$/2 
plate was used, the observations were taken through a Gunn-r
filter. To save readout time, we only used the
central 1000 $\times$ 1000 pixel of the Site-CCD, 
which gave us a field of view of 7\arcmin $\times$ 7\arcmin\
(0\farcs51/pixel). 
The layout of the observations was similar to the one
employed at the NTT, except that here only one or two polarized and
unpolarized standards from \citet{1990AJ.....99.1243T} or
\citet{1992AJ....104.1563S}
were observed during each of the nights.

We finally acquired observations of another set of targets using ALFOSC
attached to the NOT, La Palma during three clear nights with mostly
good seeing (0\farcs7 - 1\farcs5) from 
April 1 - 4, 2009. Here a calcite plate and a $\lambda$/2 
retarder plate were used. The data were taken through an 
SDSS-r'filter (NOT Nr. 84). The two beams were separated by 15\arcsec.
The observations were carried out with an E2V-CCD. Due to technical
constraints only the central 1500 $\times$ 650 pixel providing a field
of view of 4\farcm7 $\times$ 2\arcmin\ (0\farcs19/pixel) could be used.
Since both our NTT and CA observations suffered from 
substantial instrumental polarization (see section \ref{analysis}) 
we re-observed 13 sources that had been observed at the NTT and CA as a
``sanity check''. We again used the same observing layout as at the
NTT and CA. Here, one or two polarized and
unpolarized standards were observed each of the night.

The data reduction was similar in all cases. First, the images were
corrected for their bias, and the dark current was proven to be negligible in
all cases. Then we corrected for the pixel-to-pixel variations across the
CCD using either flatfields taken during twilight (NTT and NOT) or
images taken of a homogeneously illuminated screen inside the dome (CA). 

The observing log for each source
(telescope used, integration times) is given
in Table \ref{poliresults}. Table \ref{poliresults}, 
available at the CDS, contains the following information. 
Column 1 lists the J2000 coordinates of the source, Column 2 its redshift, 
Column 3 whether a source is listed in NED, and Column 4 the SDSS r-mag. 
The entries listed in Columns 2-4 are from C05. Column 5 gives the 
telescope for the observations used, Column 6 the date of the observations
and Column 7 the exposure time for an individual exposure per 
position angle. Columns 8 and 9 give the measured degree of polarization, 
as well as the position angle. Finally in Column 10 we give 
references to previous measurements of the sources.

\section{\label{analysis}Analysis}

The normalized Stokes parameters $P_Q$ and $P_U$ were computed in the
same way for all three datasets (NTT, Calar Alto, and NOT). We first
used aperture photometry to measure the fluxes in the ordinary and
extraordinary beams in each of the four positions of the Wollaston
prism/calcite. The measurements were made with aperture radii of
1\farcs3 - 3\farcs5 (but mostly between 1\farcs5 and 2\farcs0) 
 depending on seeing and object brightness.
In addition to the BL Lac candidate, these measurements
also included any sufficiently bright stars present on the CCD
frame, where ``sufficiently bright'' means that the errors of both $P_Q$ and
$P_U$ are smaller than 0.5\%. In this paper we express $P_Q$,
$P_U$ and $P$ in percentage, so a 0.5\% error does not imply an S/N
= 200. We then computed $P_Q$ and $P_U$ using standard
formulae \citep[e.g.][]{2009MNRAS.397.1893V}. In this phase, we checked
that there were no spurious objects inside the measurement aperture. 
In the few cases where such an object was detected, the contaminating flux
was measured and subtracted.  From $P_Q$ and $P_U$, the degree of
polarization $P$ and polarization position angle PA were then
calculated from $P = \sqrt{P_Q^2 + P_U^2}$ and $PA = 1/2
\tan^{-1}(P_U/P_Q)$.

The errors of $P_Q$ and $P_U$ were computed by propagating the flux
measurement errors through the formulae. Typical errors are $\sim$
0.8\% for $P_Q$, $P_U$ and $P$ and $\sim$ 4$^{\circ}$ for PA. In
addition, small systematic errors are present owing to the correction of
instrumental polarization and a mismatch between the filters used in
this study and the ones used in the literature. As discussed below, the
systematic errors are expected to be smaller that typical errors bars
($<$0.3\% in $P_Q$ and $P_U$ and $<2^{\circ}$ in PA).

Since especially NTT/EFOSC2 was expected to exhibit high
instrumental polarization, we took special care to characterize
possible instrumental effects at all three telescopes. At the NTT we
made observations of zero polarization standards in
\citet{2007ASPC..364..503F} at 62 positions on the sky to map how the
instrumental polarization depends on telescope orientation. In these
measurements the object was placed near the center of the CCD. In
addition, we mapped the instrumental polarization as a function of
position on the CCD by observing three zero polarization standards in
a 5$\times$5 grid over the whole field of view.

\begin{figure}
\epsfig{file=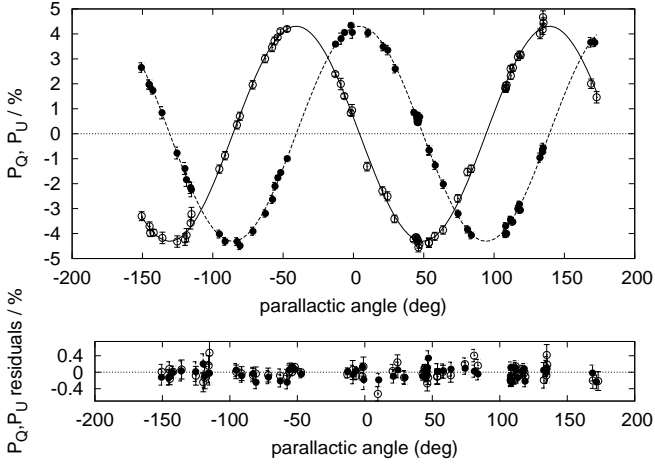,width=9cm}
\caption{\label{parangplot} {\em Upper panel}: Dependence of NTT/EFOSC2 
instrumental polarization on parallactic angle. Open and closed symbols
refer to the normalized Stokes parameters $P_Q$ and $P_U$, respectively.
The solid and dashed lines are fits to the $P_Q$ and $P_U$
data, respectively  (Eqs. \ref{ekay} and \ref{tokay}). 
{\em Lower panel}: Residuals after subtracting the fit.}
\end{figure}

Figure \ref{parangplot} shows $P_Q$ and $P_U$ as a function of
parallactic angle of the CCD as measured from zero polarization stars
at the NTT. There is a high instrumental polarization present with an
amplitude of (4.31$\pm$0.02)\%, modulated by the parallactic
angle. We fitted the functions
\begin{eqnarray}
\label{ekay}
P_Q &=& A * \cos(\theta - \theta_0)\\
\label{tokay}
P_U &=& A * \cos(\theta - \theta_0 - \pi/2)
\end{eqnarray}
to the observed data where $\theta$ is the parallactic angle and 
$A$ and $\theta_0$ are the fitting variables. The fits are shown as
in Fig. \ref{parangplot}. Subtracting this fit leaves
no significant residuals (Fig. \ref{parangplot}, lower panel).

After removing the dependence on parallactic angle, we found the
instrumental polarization to also depend on the position on the
CCD. The instrumental polarization was zero at the center of the CCD
and increased to $\sim$0.5\% in the corners. The field dependence was
modeled by fitting two-dimensional polynomials
up to second degree to $P_Q$ and $P_U$ and removed. After this no
significant residuals above the error bars (0.1-0.2\%) were seen.

\begin{figure}
\epsfig{file=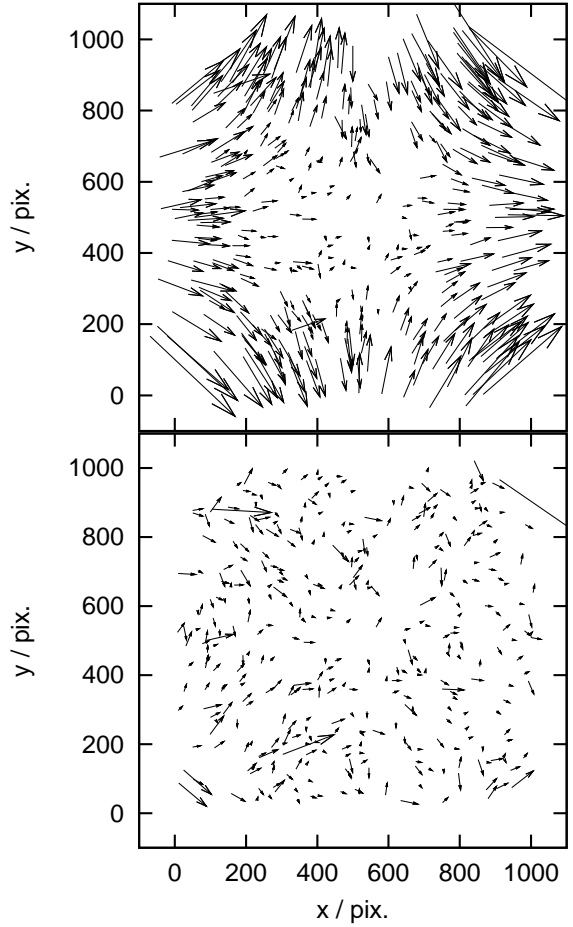,width=8cm}
\caption{\label{cafieldplot} 
Degree of polarization of field stars with the Calar Alto 2.2m
telescope/CAFOS as a function of position on the CCD before (upper
panel) and after (lower panel) applying the corrections. An arrow
with a length of 100 units corresponds to 1\% polarization.}
\end{figure}

At Calar Alto we examined the polarization of 360 field stars present
in the CCD frames.  These stars are not necessarily unpolarized, but
since they are not at low galactic latitudes, their polarization is
expected to be low, below, or close to our error bars
(0.05-0.5\%). There is a clear dependence of polarization on the
position on the CCD (Fig. \ref{cafieldplot}) reaching 3-4\% near the
corners. As with the NTT data, we fitted up to
second-order, two-dimensional polynomials to $P_Q$ and $P_U$ and
subtracted the fit from the data. After subtraction no residuals above
the rms noise in $P_Q$ and $P_U$ (0.2\%) were seen.

The NOT data show no instrumental effects above the rms noise (0.3\%)
of 26 field stars. We can thus say that any remaining instrumental
effects in our whole data set are below 0.2-0.3\% in $P_Q$ and $P_U$,
considerably less than typical error bars in BL Lac candidates
{\bf ($\sim$ 0.8\%).}

To determine whether a target is polarized, we computed the 95\%
confidence limits of the observed degree of polarization $P$ using the
formalism in \citet{1985A&A...142..100S}. If the lower confidence
limit of $P$ is $> 0$, we denote the target as polarized. In this case
the unbiased degree of polarization was computed using the maximum
likelihood estimator in \citet{1985A&A...142..100S}; i.e., $P_{\rm
  unbiased} = \sqrt{P^2-1.41*\sigma_P^2}$ where $\sigma_P =
(\sigma_{P_Q} + \sigma_{P_U})/2$. If the target was unpolarized
(i.e. the lower 95\% confidence limit = 0), we used the upper 95\%
confidence limit as the upper limit for the degree of polarization.

The polarization position angle was calibrated by making 21
observations of six highly polarized stars in
\citet{1992AJ....104.1563S} and \citet{2007ASPC..364..503F}.  We used
the quoted R-band values to determine the position angle zero
point. The derived zero points have an rms scatter of $\sim$
1$^{\circ}$ internally for each instrument.  Since we used filters
that are
slightly different from the R-band and since the polarization position
angle in high polarization standards is typically wavelength-dependent, 
though not strongly, a small systematic error in our PA
calibration was expected. We compared the derived r-band zero points to
the ones derived in \cite{2010MNRAS.402.2087V} for the R-band and
found the two to differ by 1.4 $\pm$ 1.3 and by 1.8 $\pm$ 1.1 degrees
for the CA and NOT data, respectively. As a result, any PA offsets due to
filter mismatch are likely to be smaller than 2 degrees.

\section{Results}

The results for each object individually are presented in  
Table \ref{poliresults}, while we give 
a breakdown of our results in Table \ref{breakdown} 
as discussed below.

In total, we have 195
measurements (123 NTT, 47 CA, 25 NOT) of 182 targets. According to
C05, 135 out of 182 targets were not reported in
NED before they published their sample.
Thirteen of our 182 targets were observed twice using the NOT and NTT or CA,
and 10 out of 13  were not listed in NED when C05 published their catalog. 

\setcounter{table}{1}
\begin{table*}
\caption{\label{breakdown} Statistics of the observed sample.}
\begin{tabular}{l|c|cc|cc|cc}
\hline
\hline
 & N & NED\tablefootmark{1} & not in NED & z\tablefootmark{2} &
no/uncertain z & X or r\tablefootmark{3} & no X or r\\
\hline
Polarized targets& 124 & 39 & 85 & 44 & 80 & 124 & - \\
\hline
Unpolarized targets & 58 & 8 & 50 & 23 & 35 & 39 & 19\\
\hline
\end{tabular}
\tablefoot{
\tablefoottext{1}{NED entry as BL Lac for the target quoted in C05.
\tablefoottext{2}{A reliable redshift exists for the target in}
C05.}
\tablefoottext{3}{X-ray or radio data exist for the target in the literature.}
}
\end{table*}

\begin{figure}
\epsfig{file=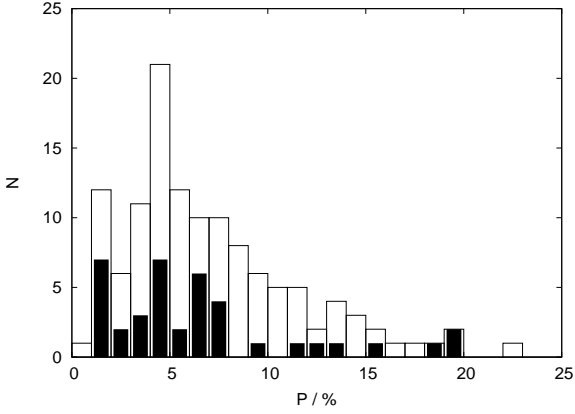,width=8cm}
\caption{\label{pdistr} 
Distribution of the degree of polarization of our targets. The
candidates with entries in NED are indicated in black.} 
\end{figure}

Out of our 182 targets 124 (68\%) have been found to be polarized and 
95 out of the 124 (77\%) polarized objects have been found to be 
 highly polarized with 
P $>$ 4\%\footnote{Following \citet{1993ApJS...85..265J} we set the
  border line to distinguish between weakly and highly polarized
  objects to 4\%.}. 
The average polarization is 7\%, with a substantial tail
of 27 targets whose polarization exceeds 10\% and a maximum
polarization of 22\%. Figure \ref{pdistr} shows the distribution of the
polarization degree of our targets. There is
basically no difference in the distribution of the ``already known''
and really new BL Lac candidates. Indeed, a K-S test confirmed that 
the null hypothesis that the two p distributions are drawn from the
same parent population cannot be rejected (significance
0.204).
We also inspected the
distribution of the polarization angles. As expected we do not find
any preferred polarization angle. 

Only 44 out of 124 (35\%) of our polarized sources have a reliable redshift,
33 have lower limits and/or uncertain redshifts while for 47 targets
no redshift is available at all. The redshifts are taken from the
catalog of C05. Their reliable redshifts are based on at least two spectral
features (mostly but not always from the host galaxy absorption
features), lower limits are derived from intervening absorption 
systems (typically $[$ Mg II $]$), while uncertain redshifts 
are based on one single 
spectral feature. 
Our results clearly indicate the need for deep follow-up spectroscopy. 

\begin{figure}
\epsfig{file=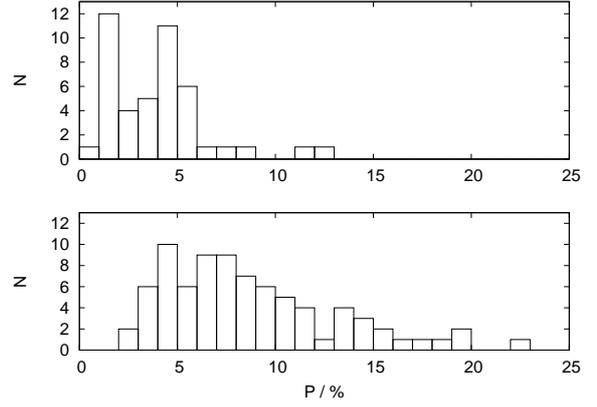,width=8cm} 
\caption{\label{zpolwoz} 
Distribution of the degree of polarization of our 44 targets with
reliable redshifts (upper panel) and of our 80 targets with either
lower limits, uncertain, or no redshifts at all (lower panel).}
\end{figure}

Figure \ref{zpolwoz} shows the distribution of the degree of
polarization for the 44 BL Lac candidates with reliable redshift and the 80
remaining ones in separate panels. Obviously, the two distributions
differ in the sense that the targets with reliable z tend to be less
polarized. The median polarization for the targets with reliable
redshift is 3.8\% and for the remaining ones 7.8\%, respectively.
A KS-test shows that the two distributions are not different at a $<$
0.01\% level. This results can best be explained by contamination of
our polarization measurements by host galaxy light for the sources
with reliable redshifts (see above and discussion in
section \ref{hbllbl}).

\begin{figure}
\epsfig{file=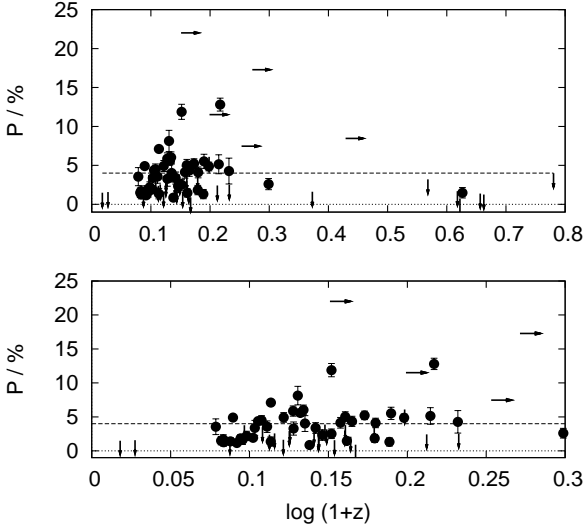,width=8cm} 
\caption{\label{z_pol} Polarization versus redshift of our targets
with reliable redshifts (black dots), lower limits to the
redshift (right arrows), and upper limits to the polarization (down
arrows). The dashed line marks 4\% polarization. The upper panel
displays the results for the full data set, while the lower panel
shows our results for redshifts up to z = 1 only.}
\end{figure}

In Fig. \ref{z_pol} we show the behavior of our sources in the
polarization - redshift plane. There seems to be a trend toward
sources with higher redshifts to be more polarized roughly up to z
$\sim$ 1. This would indicate
a bias, since at higher redshift only the more beamed sources can be
detected and the dilution by host galaxy light is much lower. (Our 
polarization measurements have not yet been corrected for host galaxy 
contamination.) 
Above z $\sim$ 1, the polarization of the sources seems to drop.  
There is one weakly polarized source at a redshift $>$ 3 and a few
more sources where we could derive only upper limits.
These may be high-redshift weak-line QSOs (see discussion in section
\ref{rqbllacs}). However, there are
potentially five more highly and two more weakly polarized sources above 
z = 1 in our sample, but their redshifts are marked as uncertain in C05.
To test for a correlation of polarization with redshift, 
we calculated the Spearman rank correlation coefficient for all
43  sources with reliable redshifts up to z = 1. We only find $\rho$ =
0.284. A Student's T-test confirmed that $\rho$ is not
significantly  different from 0 (${\tau}_{\rho}$ = 1.90). We also tested
whether the polarization properties of the 22 low-z (z $<$ 0.35) and
the 21 high-z (z $>$ 0.35) sources differ. According to a KS-test, the
polarization properties of the two subsamples are not significantly
different. The expected bias is not as pronounced as one may expect. This
could come from the small apertures for our polarimetry used (see
discussion in section \ref{hbllbl}).

All of our 124 polarized sources have a radio detection in either
FIRST
\citep[faint images of the radio sky at twenty centimeters,][]
{1995ApJ...450..559B}
or NVSS \citep[NRAO VLA Sky Survey,][]{1998AJ....115.1693C}, 
or have recently been detected in deep VLA observations by 
\citet[][P10b hereafter]{2010ApJ...721..562P}.
Only 44 (35\%) of our 124 sources are detected in both the radio and X-ray
regime. This is not surprising since the X-ray measurements
were taken from the RASS \citep[ROSAT All-Sky Survey, 
e.g.][]{1993Sci...260.1769T}, which is generally much shallover 
than the radio surveys used for cross-correlation. Upcoming deep X-ray
surveys like the one planned by eRosita will certainly detect a major
fraction of the sources discussed here. Nevertheless, most
targets that have been found to be polarized are true new ``bona
fide'' BL Lac objects. Only 84 out of 240 sources were listed 
as BL Lac in NED at
the time when C05 published his sample.
In the meantime, the number of
NED entries is much larger, but this is exclusively due to newly
presented samples by A07 and P10a.

Remarkable is the breakdown of our 58 unpolarized sources. 
Thirtynine of them 
(67\%) have either a radio counterpart (25 sources, with one
recently been detected by P10b; SDSS J21155288+000115.5, but still 
radio-quiet) or a X-ray detection (1 source), or they have been detected in
radio and X-rays (13 sources). 
About 60\% (23 out of 39) of them have a reliable redshift.

The 19 unpolarized sources that have been detected neither at radio nor at
X-ray frequencies
all belong to the potential radio-weak BL Lac candidates
presented by C05. Reliable redshifts are available for 42\% (8 out of 19) of
these sources. 

\begin{figure}
\epsfig{file=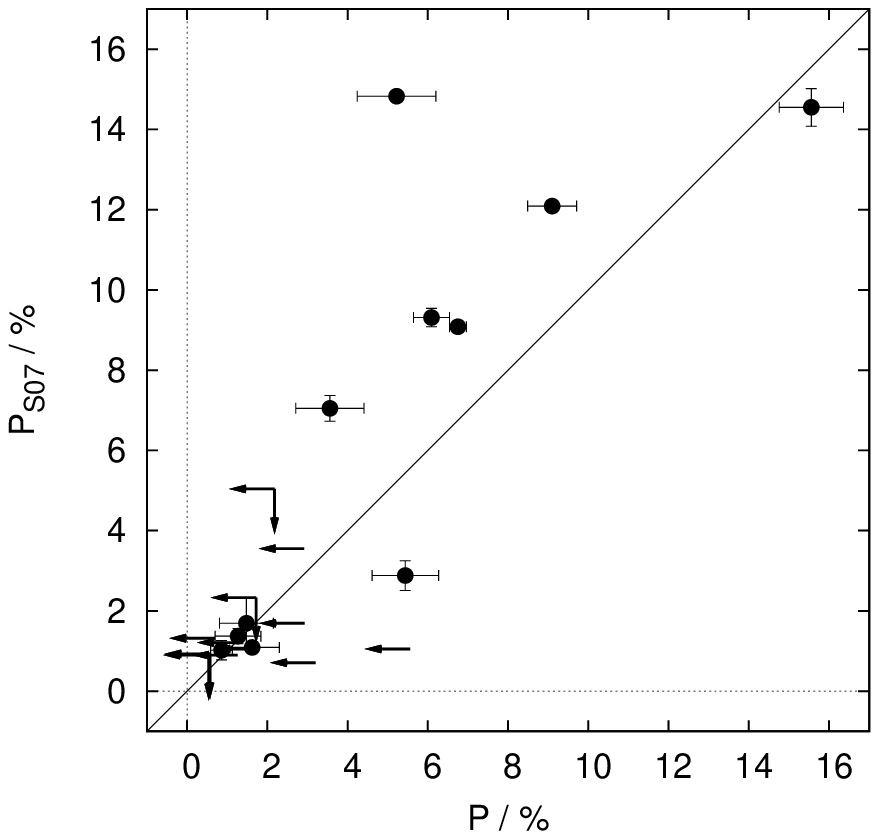,width=8cm}
\epsfig{file=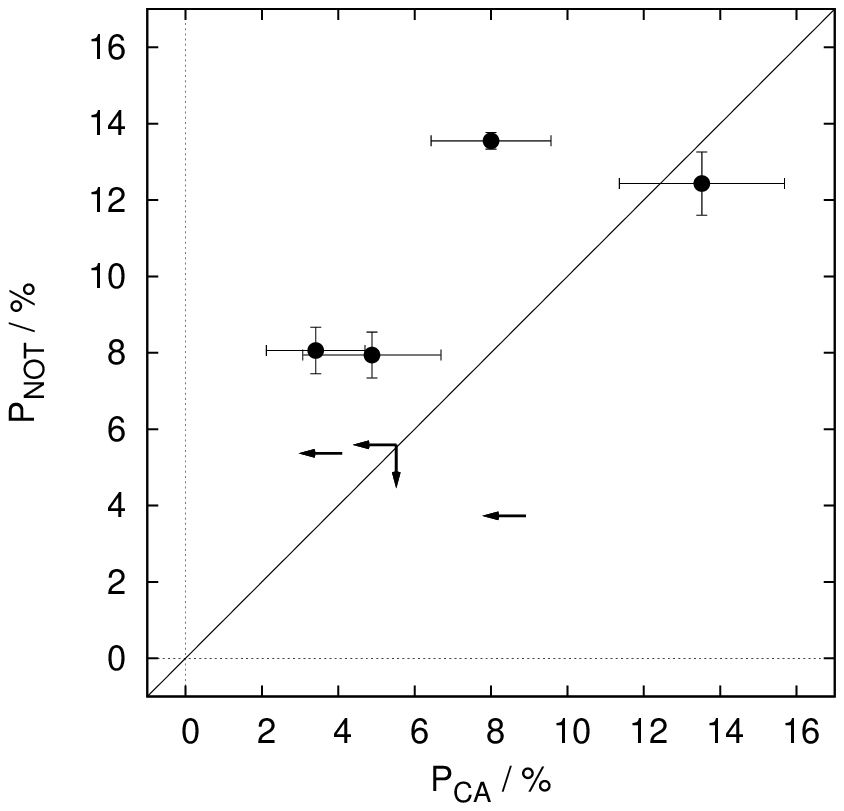,width=8cm}
\epsfig{file=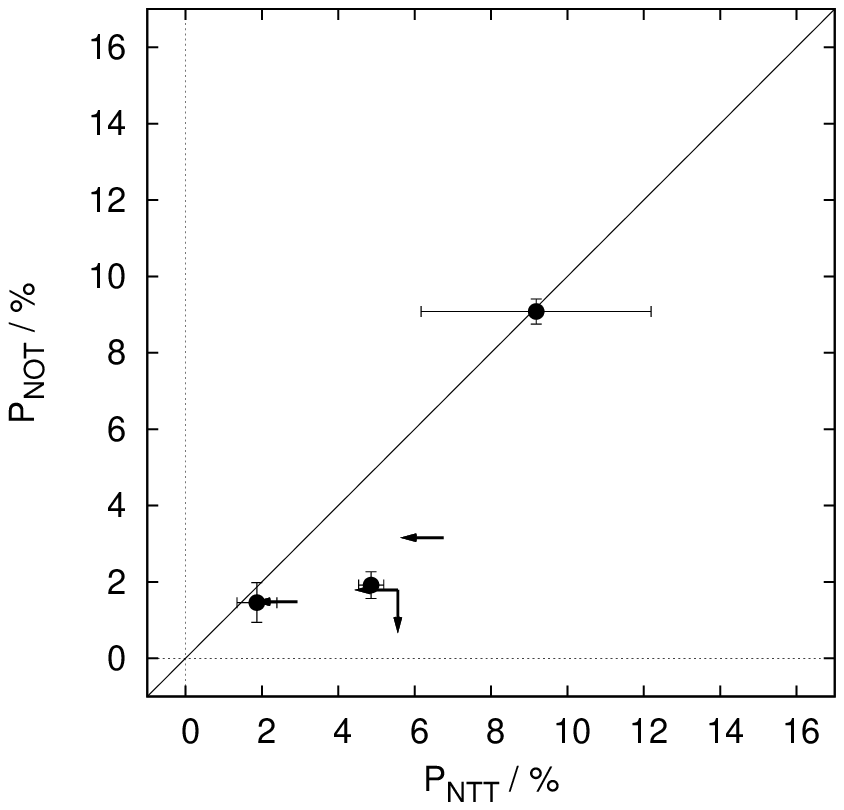,width=8cm}
\caption{\label{pol_var} 
Polarization measurements of S07 compared to ours (top), NOT vs. CA
(center) and NOT vs. NTT (bottom). The diagonal line gives the 1:1 
correspondence. The arrows indicate upper limits. Polarization
variability for a number of objects is apparent in particular at larger 
baselines.
}
\end{figure}

Our data set allows us to inspect some of our targets for 
polarization variability. Six targets have been observed with a
separation of a couple of days (NTT and NOT), seven targets with a
separation of about six weeks (CA and NOT), and 21 more targets with a
separation of three to four years (our data and S07). A comparison
of the polarization measurements taken at different epochs is
provided in Fig. \ref{pol_var}. As can be seen, a large fraction 
($\sim$ 1/3) of our BL Lac candidates show polarization variability 
indeed. Polarization measurements are available in
the literature for six more of our sources. 
These are the two 1Jy BL Lacs SDSS J005041.31-092905.1 and
SDSS J014125.83-092843.7 observed by \citet{1986MNRAS.221..739B} and 
\citet{1990A&AS...83..183M},
the two EMSS BL Lacs SDSS J020106.18+003400.2
and SDSS J140450.91+040202.2 observed by \citet{1994ApJ...428..130J},
SDSS J105829.62+013358.8 observed by \citet{1988ApJ...333..666I},
and SDSS J121834.93-011954.3 observed by
\citet{2005A&A...433..757S}. For all sources polarization between
5\% and 30\% was detected. Comparing the measurements to ours, all
six sources have shown polarization variability on timescales of years.

\section{Discussion and conclusions}

\subsection{General comments}

Although BL Lac objects are by definition polarized, have a duty cycle
of at least 40\% to be highly polarized 
\citep{1993ApJS...85..265J,1994ApJ...428..130J}, and their polarization
properties can be used to probe radiation processes in their central
engines
\citep[e.g.][]{2008Natur.452..851V,2010MNRAS.402.2087V,2010Natur.463..919F},
polarization measurements have rarely been used to verify them.
So far, only \citet{1982MNRAS.201..849I},
\citet{1984ApJ...276..449B},
\citet{1988ApJ...333..666I},
\citet{1988A&AS...76..145F}, 
\citet{1990AJ.....99....1K},
\citet{1993ApJ...404..100J},
\citet{1996A&A...307..745K},
\citet{1996MNRAS.281..425M}, and
S07 used polarization measurements
as a diagnostic tool for detecting or confirming BL Lac candidates.

About 124 of the 182 targets observed by us were found to be
polarized, and 95 were highly polarized. S07 observed 21 more sources
not observed by us. All of them were found to be polarized, 15 out of 21
were highly polarized. Since S07 and we concentrated on sources without
an entry in NED, it is not a surprise that the majority of the
remaining sources are already known BL Lacs, e.g., from the RGB sample
by \citet{1999ApJ...525..127L}. Polarization
measurements are published for only eight more 
sources\footnote{
SDSS J081815.99+422245.2 \citep{1990AJ.....99....1K},
SDSS J083223.22+491321.0 \citep{1990AJ.....99....1K},
SDSS J105837.74+562811.2 \citep{1996MNRAS.281..425M},
SDSS J123131.40+641418.2 \citep{1993ApJS...85..265J},
SDSS J123739.07+625842.8 \citep{1993ApJS...85..265J},
SDSS J140923.50+593940.7 \citep{1993ApJS...85..265J},
SDSS J150947.97+555617.3 \citep{1996A&A...307..745K},
SDSS J164419.98+454644.4 \citep{1996A&A...307..745K}
}, 
all of which were found to be polarized and five to be 
highly polarized.
In sum, polarization measurements are available for 211 of 240 targets (88\%)
from the C05 sample of probable BL Lac candidates, 153 (64\%) of them
were found to be polarized at least once, and 115 (48\%) 
to be highly polarized.

\subsection{\label{hbllbl}HBL versus LBL}

\begin{figure}
\epsfig{file=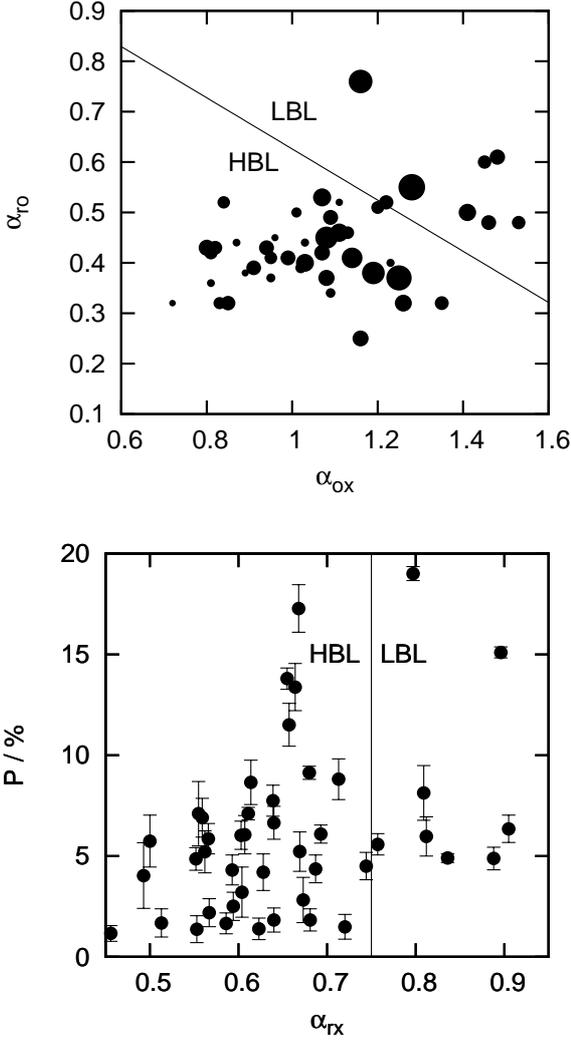,width=8cm}
\caption{\label{alphaplot}  
Upper panel: Positions of our sources with polarization in the 
$\alpha_{ox} - \alpha_{ro}$ plane. The area of the symbol is 
proportional to the degree of polarization.
The division line between HBL and LBL is also indicated. 
Only the 45 targets with radio and X-ray counterparts are 
included here. Lower panel: the degree
of polarizarion P as a function of $\alpha_{rx}$.} 
\end{figure}

Figure \ref{alphaplot} shows the location of our polarized sources 
with both radio and X-ray detection 
in the $\alpha_{ox} - \alpha_{ro}$ plane as well as 
the degree of polarizarion P as a function of $\alpha_{rx}$.
As in C05, we use $\alpha_{rx}$ = 0.75 as the dividing line between
LBL and HBL.
Obviously, more HBL than LBL are in this figure, which may be because 
the radio surveys (NVSS, FIRST) are 
much deeper than the X-ray survey (RASS).
Although we are limited by low number statistics here, it seems that the 
HBL are only slightly less polarized on average than LBLs.
The average polarization for our 
eight LBL in the diagram is 8.7 $\pm$ 1.9\% (mode 6.2\%), while it is 
5.6 $\pm$ 0.6\% (mode 5.0\%) for the 37 HBL. As ``errors'' we quote
the standard error of the mean. A K-S test confirmed that our LBL and
HBL do not show a significantly different polarization (significance
0.0758).

The two most highly 
polarized LBL are the 1 Jy BL Lac SDSS J005041.31-092905.1 and 
SDSS J105829.62+013358.8, both of which were found to be highly polarized
before. On the other hand, a number of strongly ($> 10\%$) polarized sources 
are HBL, most of them without any entry in NED. 
The object with the highest polarization in our entire data set
(22\%, SDSS J121348.81+642520.2), does not enter the plot here since
only upper limits to its X-ray flux are available. 
With $\alpha_{ox} > 1.07
$ and $ \alpha_{ro} = 0.43$ it is very much in the center of the HBL
region. To move it into the LBL region ($\alpha_{ox} \sim 1.4 $ for 
$ \alpha_{ro} = 0.43$), its X-ray flux must be a factor 7 below the
upper limit in C05. 

\begin{figure}
\epsfig{file=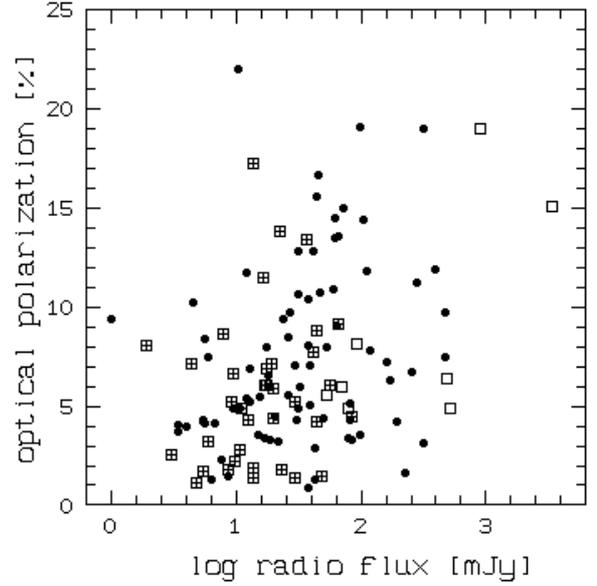,width=8cm}
\caption{\label{radiopol} Radio flux versus optical polarization 
of our SDSS BL Lac candidates. Open squares denote HBL, crossed
squares HBL, and dots candidates, where only upper limits to
X-ray fluxes exist. Only a weak correlation is apparent. }
\end{figure}

Interestingly, 25 of the 37 (68\%) HBL that enter Fig. \ref{alphaplot} were 
found to 
be $> 4\%$ polarized (and all eight LBL). This seems to be different
from the  results obtained 
by \citet{1994ApJ...428..130J}, who find a duty cycle
(fraction of time spent with P $>$ 4\% and assuming that the temporal 
distribution of polarization is equivalent to the distribution of
polarization measured in all objects of its class)
for HBL to be only about 44\%. Since LBL are stronger
radio emitters than HBL, one would not
expect any strong correlation between optical polarization and radio flux.
As Fig. \ref{radiopol} shows, this is indeed not the case. 
The Spearman rank-order correlation coefficient $r_s$
= 0.31; i.e., the two sets of data only show a weak positive correlation.
On the other hand, if we compute an ``error'' for the
duty cycle such that we count the number of of objects that could,
within their 1 $\sigma$ errors, move above and below the 4\% border
as in \citet{1994ApJ...428..130J}, we find a duty cycle for our HBL
of $68^{+2}_{-14}$\%. We note that the average redshift of our 
15 HBL with reliable redshift is z = 0.37$\pm$0.17, which 
is very similar to the ones  obtained for the EMSS or Exosat samples;
i.e., there are good reasons to assume that we are dealing with the
same class of HBL.

\begin{figure}
\epsfig{file=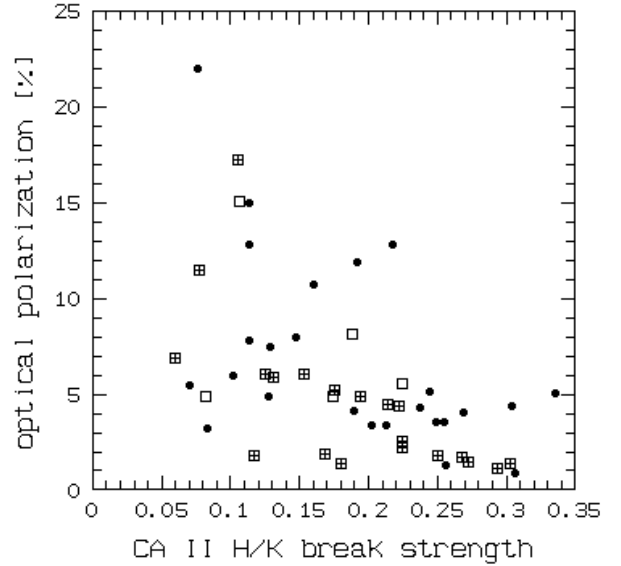,width=8cm}
\caption{\label{polihk} 
Ca II H/K break strength versus optical polarization for 50 of
  our BL Lac candidates. The symbols represent the same type of
  targets as in Fig. \ref{radiopol}. The dependence of the degree of
  polarization on break strength is obvious. HBLs are
  distributed across the entire break strength range.}
\end{figure}

Even within the ``error'', we find a higher duty cycle than  
\citet{1994ApJ...428..130J}. 
There could be several reasons for that. 
It might be simply a chance coincidence, but could also be
an effect of large errors in the derivation of the spectral radio - 
optical and optical - X-ray indices because the data are not taken 
simultaneously.
Most likely, however, it stems from a combination of 
the choice of the aperture for measuring the polarimetric fluxes
and of the seeing effects.
\citet{1994ApJ...428..130J} modeled the change in the measured
polarization with respect to the intrinsic polarization as a function
of the Ca II H/K break strength in the spectra of BL Lac objects. For
a reasonable range of break strengths, he predicted a ``depolarization''
of up to 20\%. In Fig. \ref{polihk} we show the Ca II H/K break
strength versus optical polarization for the 50 targets of our sample
(5 LBL, 25 HBL and 20 targets where only upper limits to their X-ray fluxes
are available) with the break strengths published in P10a. Obviously,
highly polarized sources (P $>$ 4\%) can be detected at all break
strengths, but the lower the break strength, the higher the
measured polarization can be. The diagram
basically confirms the prediction by \citet{1994ApJ...428..130J}. 
It shows also that HBL do not necessarily populate the righthand side of
the diagram.

\citet{1991AJ....101.1196C}, \citet{2000AJ....119.1534C}, and
\citet{2007A&A...475..199N} have shown that the choice of the aperture
may have a dramatic effect on variability measurements of BL
Lac objects due to the presence of the host galaxy. In addition,
varying seeing while the apterture is kept constant adds further
uncertainties. For their observations \citet{1994ApJ...428..130J}
used mostly a two-holer polarimeter/photometer with aperture
diamaters of 5\arcsec or larger \citep[see Table 4 in][]{1993ApJS...85..265J}. 
For the majority of our sources, we instead used apertures with a diameter 
of 4\arcsec or smaller (see section
\ref{analysis}). Although our procedure does not completely
remove the host galaxy light (see Fig \ref{polihk}), we expect that the
contamination by stellar light is much less in our case. Thus one
could expect a higher duty cycle for HBL as the one found by 
\citet{1994ApJ...428..130J}. 

To test this, we redid our polarization analysis with a fixed aperture 
diameter of 5\arcsec similar to the one used by
\citet{1994ApJ...428..130J}. Compared to our previous analysis, 
the general result did not change much. Now 116 instead 
124 out of 182 sources are polarized (64\% instead of 68\%), while 88
out of 116 
(76\%) instead of 95 out of 124 (77\%) are highly polarized. Some stronger 
differences can be seen when we compare the polarization properties of 
our 8 LBL and 36 HBL. Even with a larger aperture, all LBL 
are highly polarized. However, of 36 only 29 HBL are polarized and 22 (61\%) 
are highly polarized. When we again compute an ``error'' for the duty cycle 
as described above, we now find a duty cycle for our HBL of
$61^{+6}_{-14}$\%. This is within the ``error'' very close to the value 
derived by \citet{1994ApJ...428..130J}. We are
right now in the process of deblending the polarization measurements 
of our BL Lac candidates into the contribution of the AGN and host
galaxy. Potentially, the polarimetry of ``host galaxy free'' LBL and
HBL will show that their duty cycles do not differ. This will be 
included in a forthcoming paper.

\subsection{\label{rqbllacs}Radio-quiet BL Lacs}

The existence of radio-quiet BL Lac objects has a long and
controversial history. While e.g. \citet{1990ApJ...348..141S} did
not found any in the sample of EMSS BL Lac objects,
\citet{2004MNRAS.352..903L} potentially found a candidate from the 2QZ
BL Lac surcey \citep{2002MNRAS.334..941L}. However, as 
P10b argued, the investigated samples were
simply too small to find subsets of rarely radio-quiet BL Lac objects. 
Alternatively, radio-quiet BL Lac candidates could be low-redshift
counterparts of the (also rare) weak-line QSOs (WLQ), first detected by 
\citet{1999ApJ...526L..57F}. These objects have several properties
similar to BL Lac objects, but they do not show strong radio emission and show
very low variability and polarization
\citep[][]{2009ApJ...699..782D}. 
They apparently have an
intrinsically weak or absent broad emission line region rather than
being diluted by the beamed emission from a relativistic jet
\citep[e.g.][]{2009ApJ...696..580S,2010ApJ...722L.152S}.

\begin{table}
\caption{\label{rqtab} Properties of radio-quiet BL Lac candidates}
\begin{tabular}{l|c|c|c}
\hline\hline
SDSS J             & P   & P (S07) & P10b\\
	           & [\%]& [\%]&      \\	    	  
\hline
004054.65-091526.8 & $<$ 4.0&     &       \\
012155.87-102037.2 & $<$ 0.9&     & Unknown     \\
013408.95+003102.5 & $<$ 2.0&     & Star     \\
020137.66+002535.1 & $<$ 2.2& $<$ 5.0   & Star     \\
024156.38+004351.6 & 2.6 &     & Unknown     \\
024157.37+000944.1 & 4.0 &     & BL Lac\\
025046.48-005449.0 & $<$ 3.0&     & Star     \\
025612.47-001057.8 & $<$ 2.4&     & Unknown?    \\
031712.23-075850.4 & $<$ 3.2& (0.7)\tablefootmark{a}   &       \\
090133.43+031412.5 & $<$ 2.1&     & Unknown     \\
104833.57+620305.0 & $<$ 1.7& $<$ 2.3   & Unknown?    \\
114153.35+021924.4 & $<$ 1.3& (0.9)   &       \\
121221.56+534128.0 & $<$ 0.7& (1.3)   &       \\
123743.09+630144.9 & $<$ 1.4& (1.2)   &       \\
124225.39+642919.1 & $<$ 1.5&     & Galaxy     \\
133219.65+622715.9 & $<$ 1.7&     &       \\
142505.61+035336.2 & $<$ 0.6& $<$ 0.9   &       \\
150818.97+563611.2 & $<$ 3.7&     & Unknown     \\
151115.49+563715.4 & $<$ 5.1&     & Absorbed AGN   \\
154515.78+003235.2 & $<$ 5.6& (1.1)   & Unknown     \\
165806.77+611858.9 & $<$ 5.6& 1.0 & Absorbed AGN   \\
211552.88+000115.5 & $<$ 2.2&     & WLQ   \\
212019.13-075638.4 & $<$ 3.2&     & Star?    \\
213950.32+104749.6 & $<$ 2.4&     & Galaxy     \\
224749.55+134248.2 &  	 & 0.8 & WLQ   \\ 
231000.81-000516.3 & $<$ 3.2	 &     & Unknown?    \\
232428.43+144324.4 &     & 0.9 & WLQ   \\
\hline
\end{tabular}
\tablefoot{
\tablefoottext{a}{
The measurements of targets polarized according 
to S07 but unpolarized
according to our 95\% criterion are given
in parenthesis.
}
}
\end{table}

C05 extracted a set of 27 radio-quiet BL Lac candidates out of his
probable sample of 240 BL Lac objects (a new set of 86 weak-featured
radio-quiet objects has more recently been presented by
P10a). We have polarization measurements for
25 out of 27 of them, which are summarized in Table \ref{rqtab}.
For comparison, the results from S07 are given using our 95\% criterion
of whether a source is polarized or not and from P10b, who presented 
new target identifications.

All except SDSS J024156.38+004351.6 (weakly polarized
with P = 2.6\%) and SDSS J024157.37+000944.1 (highly polarized with P =
4.0\%, respectively) were found to be unpolarized by us. 
S07 observed 11 out of 27 candidates and
found that three are weakly polarized (SDSS J165806.77+611858.9 with P =
1.0\%, SDSS J224749.55+134248.2 with P = 0.8\% and SDSS
J232428.43+144324.4 with P = 0.9\%, respectively) according to our 
95 \% criterion. S07 and we have nine targets in common. 
Except SDSS J165806.77+611858.9, which S07 found to be
polarized, while we could only derive upper limits, 
the remaining ones did not show any polarization in
either case. Altogether S07 and we observed all 27 radio-quiet BL
Lac candidates and found four to be weakly polarized and one to be
highly polarized. Recently, P10b has 
examined 20 candidates from C05 via variability measurements using
SDSS data, new spectral classifications, proper motions, and
new radio and X-ray measurements. They classified three
targets as low-redshift WLQ, two as absorbed AGNs, four as stars, two
as galaxies, eight as unknown, and only one as radio-loud BL Lac. 
Our observations confirm that the number of reliable radio-quiet BL Lac
objects in the C05 sample must be low. The only highly polarized
radio-quiet BL Lac candidate 
(SDSS J024157.37+000944.1) was classified as
radio-loud BL Lac by P10b based on new VLA
measurements, while the remaining four weakly polarized radio-quiet BL
Lac candidates were classfied as unknown,
absorbed AGN or WLQ. 
As already stated by P10b and a few times in the
present paper, besides the lack of
reliable spectroscopic redshifts for many targets, deep X-ray
observations would be very useful for properly evaluating the contents 
of the C05 sample.

\subsection{Final comments}

In summary, we have found that 

$\bullet$ 124 out of 182 (68\%) of our targets were polarized, and 95 out
of the 124 polarized targets (77\%) to be highly polarized ($>$ 4\%).

$\bullet$ Only 44 out of 124 (35\%) of our polarized 
sources have a reliable redshift. There is 
a clear need for follow-up high S/N spectroscopy.

$\bullet$ LBL are on average only slighly more strongly 
polarized than HBL. 
We found a higher duty cycle 
of polarization in HBL ($ \sim 66\%$ have polarization $> 4\%$) 
than in \citet{1994ApJ...428..130J}. This may be due to the different
apertures used in the analysis.

$\bullet$ Our data do not give any evidence for the presence 
of radio-quiet BL Lac objects in the sample of C05. 

By just using our (and S07)  polarization 
measurements, we find strong evidence that the sample selected 
by C05 indeed contains a large number of bona fide BL Lacs. At least 
70\% of the sources were found to be polarized. Since even very prominent 
BL Lacs like OJ 287 are unpolarized from time to time 
\citep[e.g.][]{2010MNRAS.402.2087V}, the number of bona fide BL Lac objects 
in the sample is presumably even higher. We have further 
options for testing this result. First of all, BL Lac objects are variable, with 
LBL and HBL having duty cycles of $\sim$ 40 and 80\%, respectively 
\citep{1996A&A...305...42H,1998A&A...329..853H}. 
Since our data were taken through filters similar to the one
used by the SDSS, we can look for variability in our targets on timescales of 
years. In addition, we can use our data to analyze the images for the 
presence of a core repesenting an AGN and a host galaxy. This would 
demonstrate that these targets where a host galaxy has been found 
are extragalactic in nature. Finally, we can use the SDSS and UKIDSS
to construct broad-band SEDs of our targets and to elegantly
identify stars, which may still contaminate the sample,
by their blackbody continuum radiation. All three of the above are in progress. 
Along with our polarization analysis, we have four diagnostic tools at
hand to identify the BL Lac content of the 
sample of C05. This will be the scope of a forthcoming paper
(Nilsson et al., in prep.). 

Like all BL Lac samples, the C05 sample seriously suffers from the 
lack of reliable redshifts, which are a prerequisite  for the construction
of a lumnosity function of BL Lacs, among other possibilities. 
It is thus no surprise that 
BL Lac luminosity functions still suffers from low number statistics
\citep[e.g.][]{2007ApJ...662..182P}. The C05 targets 
were selected with
the requirement of having S/N $>$ 100 over at least one
of three 500 \AA\ wide spectral bands in the SDSS spectra. 
The S/N per resolution
element is thus much lower than 100 for many targets.
\citet{2006AJ....132....1S} has
shown that S/N $>$ 100 per resolution element is required for
detecting the very faint emission lines and/or host galaxy absorption
features. We are now in the process of collecting high S/N spectra for all
C05 sources that we found to be polarized and whose redshift is
uncertain or unknown. 
In combination with the results from our analysis, we will
then be able to derive the necessary steps towards constructing
the first luminosity function of an optically selected sample of
BL Lac objects.

\begin{acknowledgements} We would like to thank the anonymous referee 
for constructive and detailed suggestions, that improved the 
presentation of 
our results. We would also like to thank the staff at Calar Alto 
for collecting the excellent data in Service Mode, as well as the staff at the NTT 
and the NOT for their superb support during the observations. JH acknowledges 
support by the Deutsche Forschungsgemeinschaft  (DFG) through grant HE 2712/4-1.
\end{acknowledgements}

\bibliographystyle{aa}

\bibliography{16541}

\onltab{1}{
\begin{longtable}{lcccccrccl}
\caption{\label{poliresults}Observing log and polarization results.}\\
\hline\hline
Object  & z & NED\tablefootmark{a} & rmag\tablefootmark{b} & Tel.\tablefootmark{c} & Date & Exp. & P & PA & Prev.\\
$[{\rm J}2000]$ &   &      &      &      &      & [s]  & [\%] & [deg.] & obs.\tablefootmark{d}\\
\hline 
\endfirsthead
\caption{continued.}\\
\hline\hline
Object  & z & NED\tablefootmark{a} & rmag\tablefootmark{b} & Tel.\tablefootmark{c} & Date & Exp. & P & PA & Prev.\\
$[{\rm J}2000]$ &   &      &      &      &      & [s]  & [\%] & [deg.] & obs.\tablefootmark{d}\\
\hline 
\endhead
\hline
\endfoot
000121.47-001140.3 &    0.4620  &    & 19.73 & NTT & 2008-10-04 &  360 &  4.34$\pm$0.75 &  159.2$\pm$3.1 &    \\
002142.26-090044.4 &    0.6481  &    & 19.32 & NTT & 2008-10-05 &  180 & 12.80$\pm$0.83 &  127.1$\pm$1.9 &    \\
002200.95+000658.0 &    0.3057  &  Y & 19.28 & NTT & 2008-10-05 &  180 &  $<$2.56       &                &    \\ 
002839.77+003542.2 &    0.6866? &    & 19.63 & NTT & 2008-10-04 &  360 &  4.45$\pm$0.69 &   64.7$\pm$3.0 &    \\
003514.72+151504.1 &            &  Y & 16.59 & NTT & 2008-10-06 &   20 &  6.90$\pm$0.96 &   83.5$\pm$2.8 &    \\
004054.65-091526.8 &    5.0300  &    & 20.50 & NTT & 2008-10-04 &  600 &  $<$3.99       &                &    \\ 
005041.31-092905.1 &            &  Y & 16.03 & NTT & 2008-10-06 &   10 & 19.01$\pm$0.35 &   85.2$\pm$1.1 &    \\
010058.19-005547.8 & $>$0.6679? &  Y & 19.21 & NTT & 2008-10-06 &  180 & 12.82$\pm$0.66 &   12.5$\pm$1.7 &    \\
010326.01+152624.8 &    0.2461  &    & 18.07 & NTT & 2008-10-06 &   60 &  1.62$\pm$0.68 &   58.7$\pm$4.6 & S07\\
011012.66-004746.9 &    0.5477  &    & 20.04 & NTT & 2008-10-04 &  360 &  5.52$\pm$0.90 &  112.3$\pm$3.0 &    \\
011452.77+132537.5 &            &    & 17.03 & NTT & 2008-10-06 &   60 &  9.10$\pm$0.61 &  170.0$\pm$2.0 & S07\\
012155.87-102037.2 &    0.4695  &    & 19.52 & NTT & 2008-10-05 &  360 &  $<$0.90       &                & P10b\\ 
012227.38+151023.1 &            &    & 19.55 & NTT & 2008-10-06 &  360 & 10.41$\pm$1.05 &  123.1$\pm$2.4 &    \\
012716.31-082128.9 &    0.3620? &  Y & 19.14 & NTT & 2008-10-06 &  180 &  6.32$\pm$0.87 &  164.0$\pm$2.8 &    \\
012750.83-001346.6 &    0.4376  &    & 19.83 & NTT & 2008-10-04 &  360 &  4.13$\pm$0.78 &  128.1$\pm$3.3 &    \\
013408.95+003102.5 &            &    & 19.90 & NTT & 2008-10-05 &  360 &  $<$2.03       &                & P10b\\ 
014125.83-092843.7 & $>$0.5000? &  Y & 17.21 & NTT & 2008-10-06 &   90 &  4.88$\pm$0.56 &   97.3$\pm$2.6 &    \\
020106.18+003400.2 &    0.2985  &  Y & 18.25 & NTT & 2008-10-06 &   90 &  1.37$\pm$0.67 &   90.8$\pm$4.8 &    \\
020137.66+002535.1 &            &    & 19.54 & NTT & 2008-10-04 &  360 &  $<$2.18       &                & S07,P10b\\ 
022048.46-084250.4 &    0.5252? &    & 18.27 & NTT & 2008-10-06 &   90 &  6.09$\pm$0.45 &   94.3$\pm$2.1 & S07\\
023813.68-092431.4 &    0.4188  &  Y & 19.63 & NTT & 2008-10-06 &  480 &  2.51$\pm$0.70 &   25.2$\pm$3.9 &    \\
024156.38+004351.6 &    0.9900  &    & 19.61 & NTT & 2008-10-06 &  360 &  2.60$\pm$0.72 &   56.2$\pm$3.9 & P10b\\
024157.37+000944.1 &    0.7896? &    & 20.56 & NTT & 2008-10-04 &  600 &  4.03$\pm$1.63 &   81.3$\pm$4.5 & P10b\\
024302.93+004627.3 &    0.4089  &  Y & 19.46 & NTT & 2008-10-06 &  240 &  $<$3.17       &                &    \\ 
024752.13+004106.3 &    0.3929  &    & 20.18 & NTT & 2008-10-05 &  600 &  $<$2.40       &                &    \\
025046.48-005449.0 &            &    & 20.00 & NTT & 2008-10-05 &  360 &  $<$2.95       &                & P10b\\
025612.47-001057.8 &    0.6302  &    & 20.35 & NTT & 2008-10-05 &  600 &  $<$2.42       &                & P10b\\
030235.78-075027.0 &            &    & 20.33 & NTT & 2008-10-05 &  600 &  8.42$\pm$0.71 &  176.6$\pm$2.2 &    \\
030240.30+003849.9 &            &    & 20.13 & NTT & 2008-10-05 &  600 &  $<$1.93       &                &    \\ 
030433.96-005404.7 &    0.5112  &  Y & 18.76 & NTT & 2008-10-06 &  120 &  1.83$\pm$0.60 &  129.0$\pm$4.1 &    \\
031712.23-075850.4 &    2.6993  &    & 18.82 & NTT & 2008-10-04 &   90 &  $<$3.20       &                & S07\\ 
032343.62-011146.1 &            &    & 16.81 & NTT & 2008-10-04 &   20 &  5.22$\pm$0.98 &  113.5$\pm$3.2 & S07\\
032356.64-010829.6 &    0.3923  &    & 20.04 & NTT & 2008-10-06 &  360 &  $<$2.13       &                &    \\
040911.36-055529.4 &            &    & 19.78 & NTT & 2008-10-05 &  360 &  9.76$\pm$1.33 &  154.3$\pm$2.8 &    \\
045128.96-002911.5 &            &    & 20.41 & NTT & 2008-10-05 &  600 &  $<$3.98       &                &    \\
074054.60+322601.0 & $>$0.9460? &    & 18.67 & NOT & 2009-04-02 &  400 &  5.44$\pm$0.83 &   35.7$\pm$2.9 & S07\\
075144.94+392817.6 &    0.4338? &    & 20.70 &  CA & 2009-02-18 & 1000 &  3.99$\pm$1.78 &  130.3$\pm$4.8 &    \\
075602.72+414039.8 &    0.5788  &    & 19.91 &  CA & 2009-02-18 &  720 &  $<$6.09       &                &    \\
081840.06+315348.2 &            &    & 19.20 &  CA & 2009-02-19 &  720 & 14.48$\pm$1.17 &   22.4$\pm$2.5 &    \\
083413.90+511214.7 &            &    & 19.77 &  CA & 2009-02-22 &  720 &  $<$5.61       &                &    \\
083918.75+361856.1 &    0.3343  &    & 19.60 & NOT & 2009-04-02 & 1000 &  $<$2.92       &                & S07\\
084225.52+025252.7 &    0.4251  &    & 19.04 & NTT & 2009-03-29 &  200 &  $<$1.57       &                &    \\
084908.81+020622.5 &            &    & 19.07 & NTT & 2009-03-30 &  400 & 10.94$\pm$0.71 &  170.6$\pm$1.9 &    \\
085638.50+014000.7 &    0.4479  &    & 19.33 & NTT & 2009-03-31 &  200 &  $<$3.78       &                &    \\
085749.80+013530.3 &    0.2812  &  Y & 17.91 & NTT & 2009-03-29 &   40 &  4.50$\pm$0.68 &    4.7$\pm$2.9 &    \\
085920.56+004712.1 &            &  Y & 18.76 & NTT & 2009-04-01 &  200 &  4.20$\pm$0.91 &   36.1$\pm$3.5 &    \\
090133.43+031412.5 &    0.4591  &    & 18.96 & NTT & 2009-03-30 &  200 &  $<$2.08       &                & P10b\\
090939.84+020005.3 &            &  Y & 19.52 & NTT & 2009-03-29 &  400 & 18.99$\pm$0.49 &  109.9$\pm$1.3 &    \\
091848.57+021321.8 &            &    & 18.55 & NTT & 2009-04-01 &  400 &  4.29$\pm$1.16 &   96.7$\pm$3.8 &    \\
092542.87+595816.3 &            &    & 19.12 &  CA & 2009-02-22 &  720 &  8.65$\pm$1.10 &   83.3$\pm$2.9 &    \\
092638.88+541126.7 &    0.8500? &    & 19.41 &  CA & 2009-02-22 &  720 &  7.02$\pm$0.93 &   24.9$\pm$3.0 &    \\
092912.25+030029.9 &            &    & 20.09 & NTT & 2009-03-31 &  600 &  9.41$\pm$0.69 &   91.4$\pm$2.1 &    \\
094245.30+541620.4 &            &    & 20.22 &  CA & 2009-02-19 & 1000 &  7.51$\pm$2.17 &   46.6$\pm$4.1 &    \\
094257.81-004705.2 &    1.3600  &    & 18.78 & NTT & 2009-03-30 &  200 &  $<$1.62       &                &    \\
094432.33+573536.2 &            &    & 19.87 &  CA & 2009-02-22 &  720 &  3.41$\pm$1.29 &  152.9$\pm$4.5 &    \\
                   &            &    &       & NOT & 2009-04-02 & 1000 &  8.06$\pm$0.61 &  168.0$\pm$2.1 &    \\
094441.48+555753.1 &            &    & 19.89 &  CA & 2009-02-19 &  720 &  $<$4.48       &                &    \\
094542.24+575747.7 &    0.2289  &  Y & 17.49 & NOT & 2009-04-04 &  500 &  4.90$\pm$0.22 &   83.6$\pm$1.6 &    \\
094620.21+010452.1 &    0.5775  &  Y & 19.91 & NTT & 2009-03-30 &  800 &  4.86$\pm$0.56 &   57.9$\pm$2.6 &    \\
095127.82+010210.2 &            &  Y & 19.23 & NTT & 2009-04-01 &  400 &  4.36$\pm$0.69 &  105.7$\pm$3.0 &    \\
095649.53+015601.8 &            &  Y & 20.11 & NTT & 2009-03-29 &  500 &  7.10$\pm$1.60 &  136.8$\pm$3.5 &    \\
100050.22+574609.1 &    0.6392  &    & 19.48 &  CA & 2009-02-22 &  720 &  5.14$\pm$1.22 &  123.6$\pm$3.8 &    \\
100326.63+020455.7 &            &    & 19.49 & NTT & 2009-03-31 &  200 & 10.21$\pm$1.07 &   32.4$\pm$2.5 &    \\
100612.23+644011.6 &            &    & 18.87 &  CA & 2009-02-20 &  300 &  8.81$\pm$1.01 &  129.3$\pm$2.8 &    \\
100959.63+014533.8 & $>$1.0900? &    & 19.60 & NTT & 2009-03-30 &  800 & 14.98$\pm$1.35 &  177.9$\pm$2.3 &    \\
101115.64+010642.7 &    0.8615? &  Y & 19.24 & NTT & 2009-04-01 &  400 &  $<$1.64       &                &    \\
101858.55+591127.8 &            &  Y & 17.75 & NOT & 2009-04-04 &  300 & 19.08$\pm$0.46 &    3.6$\pm$1.2 &    \\
101950.87+632001.6 &            &    & 18.36 &  CA & 2009-02-23 &  300 &  7.80$\pm$0.65 &   46.9$\pm$2.5 &    \\
102013.78+625010.1 &    0.2495  &    & 18.59 &  CA & 2009-02-23 &  300 &  $<$3.77       &                &    \\
102243.73-011302.5 &            &  Y & 17.33 & NTT & 2009-03-29 &   30 &  $<$2.56       &                &    \\
102523.04+040229.0 &    0.2078  &    & 18.46 & NTT & 2009-03-30 &  120 &  1.49$\pm$0.61 &   86.1$\pm$4.5 &    \\
102724.97+631753.1 & $>$0.5816  &    & 18.61 &  CA & 2009-02-23 &  300 & 11.51$\pm$1.06 &  144.0$\pm$2.6 &    \\
103208.36+040157.0 &            &    & 19.60 & NTT & 2009-04-01 & 1200 & 11.84$\pm$0.49 &  102.2$\pm$1.6 &    \\
103220.29+030949.2 &    0.3233  &    & 18.26 & NTT & 2009-03-31 &  120 &  4.92$\pm$0.71 &  133.4$\pm$2.9 &    \\
103239.07+662323.3 &            &    & 19.17 &  CA & 2009-02-20 &  720 &  3.26$\pm$1.09 &  125.2$\pm$4.3 &    \\
103940.70+053609.3 &    0.5103  &    & 19.15 & NTT & 2009-03-29 &  200 &  $<$4.01       &                &    \\
104523.86+015722.1 &            &    & 19.06 & NTT & 2009-03-30 &  400 &  $<$1.61       &                &    \\
104833.57+620305.0 &            &    & 19.85 & NOT & 2009-04-03 & 1000 &  $<$1.72       &                & S07,P10b\\
105151.84+010310.7 &    0.2654  &    & 19.06 & NOT & 2009-04-03 & 1000 &  1.92$\pm$0.35 &  140.7$\pm$3.2 &    \\
                   &            &    &       & NTT & 2009-03-31 &  400 &  4.86$\pm$0.33 &  127.2$\pm$2.0 &    \\
105606.62+025213.5 &    0.2360  &  Y & 18.58 & NTT & 2009-04-01 &  300 &  1.16$\pm$0.39 &  169.6$\pm$4.2 &    \\
105752.79-005908.3 &    0.4678? &  Y & 19.02 & NTT & 2009-03-29 &  200 &  $<$3.90       &                &    \\
105829.62+013358.8 &    0.8862? &  Y & 17.86 & NTT & 2009-03-31 &   80 & 15.10$\pm$0.27 &  133.5$\pm$1.1 &    \\
110356.15+002236.4 &    0.2747  &    & 18.58 & NTT & 2009-04-01 &  400 &  4.38$\pm$0.36 &   53.0$\pm$2.2 &    \\
110704.78+501037.9 &    0.7061  &    & 19.53 &  CA & 2009-02-22 &  720 &  4.27$\pm$1.66 &   28.9$\pm$4.6 &    \\
110735.92+022224.5 & $>$1.0750? &    & 18.59 & NTT & 2009-03-30 &  200 &  9.39$\pm$0.42 &   86.1$\pm$1.6 &    \\
111717.55+000633.6 &    0.4511  &  Y & 19.18 & NOT & 2009-04-03 & 1000 &  1.46$\pm$0.52 &  135.6$\pm$4.3 &    \\
                   &            &    &       & NTT & 2009-03-29 &  200 &  1.87$\pm$0.52 &  129.0$\pm$3.9 &    \\
113115.50+023450.2 & $>$0.4538? &    & 18.53 & NTT & 2009-03-31 &  200 &  $<$2.79       &                &    \\
113234.38+023740.3 &            &    & 19.22 & NTT & 2009-03-31 &  400 &  $<$1.83       &                &    \\
113245.61+003427.7 &            &  Y & 17.44 & NTT & 2009-04-01 &   60 &  6.35$\pm$0.68 &   98.6$\pm$2.5 &    \\
113523.70+660941.0 &            &    & 18.94 &  CA & 2009-02-24 &  300 & 13.52$\pm$2.16 &  147.1$\pm$3.2 &    \\
                   &            &    &       & NOT & 2009-04-02 &  500 & 12.43$\pm$0.83 &  145.2$\pm$2.0 &    \\
114153.35+021924.4 &    3.5979  &    & 18.61 & NTT & 2009-04-01 &  300 &  $<$1.26       &                & S07\\
114312.11+612210.8 &            &    & 17.93 &  CA & 2009-02-20 &   90 &  5.97$\pm$0.97 &  121.5$\pm$3.2 &    \\
114926.13+624332.5 &    0.7620? &    & 18.93 &  CA & 2009-02-24 &  300 &  4.88$\pm$1.81 &  126.5$\pm$4.5 &    \\
                   &            &    &       & NOT & 2009-04-03 & 1000 &  7.94$\pm$0.60 &  111.4$\pm$2.1 &    \\
115404.54-001009.9 &    0.2535  &  Y & 18.41 & NTT & 2009-03-29 &   60 &  2.19$\pm$0.69 &  159.8$\pm$4.1 &    \\
115548.41+613554.0 &            &    & 18.88 &  CA & 2009-02-22 &  300 &  $<$8.91       &                &    \\
                   &            &    &       & NOT & 2009-04-04 & 1000 &  3.73$\pm$1.18 &  132.2$\pm$4.1 &    \\
120303.50+603119.1 &    0.0653  &    & 16.36 &  CA & 2009-02-23 &   30 &  $<$1.57       &                &    \\
120658.03+052952.2 & $>$0.7911  &    & 19.73 & NTT & 2009-03-30 &  800 &  7.47$\pm$1.07 &   18.1$\pm$2.8 &    \\
120938.33+021017.2 &            &  Y & 19.24 & NTT & 2009-04-01 &  400 &  3.21$\pm$1.24 &  107.9$\pm$4.4 &    \\
121221.56+534128.0 &    3.1900  &    & 18.63 & NOT & 2009-04-02 &  600 &  $<$0.69       &                & S07\\
121300.80+512935.6 &    0.7957? &    & 18.45 &  CA & 2009-02-23 &  300 & 13.38$\pm$1.17 &  156.9$\pm$2.5 &    \\
121348.81+642520.2 & $>$0.4157  &    & 19.20 &  CA & 2009-02-22 &  720 & 22.01$\pm$1.83 &  126.1$\pm$2.5 &    \\
121500.80+500215.6 &            &    & 17.41 &  CA & 2009-02-21 &   90 &  8.00$\pm$1.57 &    9.4$\pm$3.5 &    \\
                   &            &    &       & NOT & 2009-04-03 &  500 & 13.55$\pm$0.22 &    1.2$\pm$1.0 &    \\
121649.97+054136.7 &            &    & 19.76 & NTT & 2009-03-31 &  800 & 16.66$\pm$0.88 &  172.5$\pm$1.8 &    \\
121758.72-002946.2 &    0.4188  &    & 17.73 & NTT & 2009-03-29 &   80 & 11.88$\pm$0.97 &   66.3$\pm$2.2 &    \\
121834.93-011954.3 &    0.5545? &  Y & 17.55 & NTT & 2009-04-01 &  120 & 11.24$\pm$0.36 &  136.7$\pm$1.4 &    \\
121944.98+044622.4 &    0.4891  &    & 17.88 & NTT & 2009-03-31 &   80 &  5.22$\pm$0.63 &   63.2$\pm$2.6 &    \\
121945.70-031424.0 &    0.2987  &  Y & 17.75 & NTT & 2009-03-30 &   80 &  7.10$\pm$0.31 &   84.5$\pm$1.6 &    \\
122012.14-000306.8 &            &  Y & 19.21 & NTT & 2009-03-29 &  400 &  6.65$\pm$0.82 &  166.5$\pm$2.7 &    \\
122300.31+515313.9 &    0.3650  &    & 19.56 &  CA & 2009-02-23 &  720 &  4.02$\pm$1.16 &   25.5$\pm$4.1 &    \\
122809.13-022136.1 &    0.3227  &  Y & 19.38 & NTT & 2009-03-31 &  400 &  $<$1.63       &                &    \\
123132.38+013814.0 &    3.2300  &    & 18.75 & NOT & 2009-04-04 & 1000 &  1.48$\pm$0.67 &  127.4$\pm$4.7 & S07\\
                   &            &    &       & NTT & 2009-04-01 &  300 &  $<$2.93       &                &    \\
123341.33-014423.7 &            &    & 18.31 & NTT & 2009-03-31 &  120 & 10.64$\pm$0.64 &   59.6$\pm$1.9 &    \\
123743.09+630144.9 &    3.5347  &    & 19.00 & NOT & 2009-04-02 &  500 &  $<$1.38       &                & S07\\
124225.39+642919.1 &    0.0424  &    & 17.11 &  CA & 2009-02-21 &   90 &  $<$1.49       &                & P10b\\
124425.30+044459.7 &    0.3999  &    & 19.44 & NTT & 2009-03-29 &  400 &  2.33$\pm$0.78 &  146.9$\pm$4.2 &    \\
124533.79+022825.2 & $>$1.0900? &    & 19.09 & NTT & 2009-03-30 &  400 &  5.95$\pm$0.50 &   60.3$\pm$2.2 &    \\
124602.52+011318.8 &    0.3864  &    & 16.96 & NTT & 2009-04-01 &  120 &  3.37$\pm$0.78 &  149.6$\pm$3.6 &    \\
124834.30+512807.8 &    0.3508  &    & 17.76 &  CA & 2009-02-24 &   90 &  8.13$\pm$1.36 &  121.5$\pm$3.3 &    \\
125032.59+021632.2 &            &    & 19.21 & NTT & 2009-03-31 &  400 &  9.70$\pm$0.98 &  101.0$\pm$2.4 &    \\
125359.32+624257.5 & $>$0.8680  &    & 18.77 &  CA & 2009-02-24 &  300 & 17.28$\pm$1.18 &   71.5$\pm$2.3 &    \\
125820.79+612045.6 &    0.2235  &    & 18.73 &  CA & 2009-02-21 &  300 &  $<$1.62       &                &    \\
131106.48+003510.0 &            &  Y & 17.86 & NTT & 2009-03-29 &   80 & 13.80$\pm$0.53 &   93.1$\pm$1.5 &    \\
131330.15+020105.9 &    0.3558  &  Y & 18.64 & NTT & 2009-03-31 &  200 &  5.58$\pm$0.53 &  157.1$\pm$2.3 &    \\
132301.01+043951.4 &    0.2244  &  Y & 18.21 & NTT & 2009-03-30 &  120 &  1.39$\pm$0.53 &  138.4$\pm$4.4 &    \\
132541.91-022810.1 &    0.8073? &    & 19.88 & NTT & 2009-04-01 &  500 &  5.21$\pm$1.04 &  108.8$\pm$3.3 &    \\
132759.76+645811.3 &    0.4468  &    & 19.26 &  CA & 2009-02-24 &  720 &  $<$2.99       &                &    \\
133105.71-002221.2 &    0.2426  &  Y & 18.93 & NTT & 2009-03-29 &  200 &  1.83$\pm$0.55 &   32.0$\pm$4.0 &    \\
133219.65+622715.9 &    3.1500  &    & 19.19 &  CA & 2009-02-24 &  720 &  $<$1.68       &                &    \\
134037.59-014847.6 &    0.5130  &    & 20.05 & NTT & 2009-03-30 &  600 &  4.10$\pm$0.70 &  148.6$\pm$3.1 &    \\
135738.70+012813.6 &    0.5640? &  Y & 17.82 & NOT & 2009-04-03 &  800 &  9.08$\pm$0.33 &  167.6$\pm$1.5 &    \\
                   &            &    &       & NTT & 2009-04-01 &   40 &  9.18$\pm$3.01 &  165.6$\pm$4.1 &    \\
140450.91+040202.2 &            &  Y & 16.31 & NTT & 2009-03-30 &   20 &  7.75$\pm$0.77 &  114.8$\pm$2.4 &    \\
141003.92+051557.7 &    0.5440  &    & 19.70 & NTT & 2009-03-29 &  800 &  1.31$\pm$0.59 &  132.9$\pm$4.7 &    \\
141004.65+020306.9 & $>$1.1150? &  Y & 18.15 & NOT & 2009-04-02 &  400 &  3.16$\pm$0.90 &    6.2$\pm$3.9 &    \\
                   &            &    &       & NTT & 2009-04-01 &  100 &  $<$6.76       &                &    \\
141030.84+610012.8 &    0.3833  &    & 19.14 &  CA & 2009-02-20 &  720 &  $<$3.10       &                &    \\
141826.33-023334.1 &            &    & 16.64 & NTT & 2009-03-31 &   40 &  2.84$\pm$0.44 &   34.8$\pm$3.0 &    \\
141927.50+044513.8 & $>$1.6850  &    & 18.18 & NTT & 2009-03-30 &  120 &  8.46$\pm$0.70 &   53.1$\pm$2.2 &    \\
142409.49+043452.1 &    0.6654? &  Y & 17.72 & NTT & 2009-03-29 &   80 &  7.23$\pm$0.45 &   28.0$\pm$1.9 &    \\
142505.61+035336.2 &    2.2476? &    & 18.72 & NTT & 2009-03-29 &  400 &  $<$0.57       &                & S07\\
142526.20-011825.8 &            &  Y & 19.09 & NTT & 2009-03-29 &  400 &  $<$4.28       &                &    \\
143657.71+563924.8 &            &  Y & 18.42 & NOT & 2009-04-04 &  400 &  6.06$\pm$0.95 &  151.7$\pm$3.0 &    \\
145111.69+580003.0 &    0.4053  &    & 19.37 &  CA & 2009-02-21 &  720 &  $<$3.86       &                &    \\
145507.44+025040.3 &            &    & 19.40 & NTT & 2009-03-29 &  400 & 14.42$\pm$0.80 &  111.1$\pm$1.8 &    \\
150006.49+012956.0 &    0.7083  &    & 19.95 & NTT & 2009-03-31 & 1000 &  $<$2.53       &                &    \\
150106.26+552750.9 &            &    & 19.90 &  CA & 2009-02-22 &  720 &  $<$4.10       &                &    \\
                   &            &    &       & NOT & 2009-04-02 & 1000 &  5.37$\pm$0.72 &    1.8$\pm$2.8 &    \\
150818.97+563611.2 &    2.0521? &    & 19.51 &  CA & 2009-02-23 &  720 &  $<$3.73       &                & P10b\\
151115.49+563715.4 &            &    & 20.05 &  CA & 2009-02-24 & 1000 &  $<$5.05       &                & P10b\\
153058.17+573625.2 &    1.0998? &    & 19.40 &  CA & 2009-02-22 &  720 &  8.07$\pm$1.11 &  108.6$\pm$3.0 &    \\
154515.78+003235.2 &    1.0114? &    & 18.82 & NOT & 2009-04-03 &  500 &  $<$1.79       &                & S07,P10b\\
                   &            &    &       & NTT & 2009-04-01 &  100 &  $<$5.56       &                &    \\
155848.38+022818.6 &            &    & 19.49 & NTT & 2009-03-31 &  500 &  6.56$\pm$0.57 &   18.2$\pm$2.2 &    \\
160339.49+500955.5 &    0.6209? &    & 19.37 &  CA & 2009-02-23 &  720 &  7.07$\pm$1.02 &   89.7$\pm$3.1 &    \\
160519.05+542059.9 &    0.2117  &  Y & 18.83 & NOT & 2009-04-04 & 1000 &  1.68$\pm$0.70 &  106.2$\pm$4.5 &    \\
161541.22+471111.8 &    0.1986  &    & 17.70 &  CA & 2009-02-23 &   90 &  3.55$\pm$1.16 &  164.7$\pm$4.3 &    \\
162115.21-003140.4 &    0.4132? &    & 18.99 & NTT & 2009-03-29 &  300 &  4.31$\pm$0.74 &  156.2$\pm$3.1 &    \\
162259.24+440142.9 &            &    & 18.85 &  CA & 2009-02-21 &  300 &  $<$1.58       &                &    \\
165109.18+421253.5 &    0.2686  &    & 19.36 &  CA & 2009-02-24 &  720 &  3.38$\pm$1.10 &   87.8$\pm$4.3 &    \\
165248.44+363212.6 &    0.6470? &    & 19.22 &  CA & 2009-02-25 &  720 &  8.01$\pm$0.89 &   86.9$\pm$2.8 &    \\
165806.77+611858.9 & $>$1.4100? &    & 20.70 &  CA & 2009-02-25 & 1000 &  $<$5.52       &                & P10b\\
                   &            &    &       & NOT & 2009-04-03 & 1000 &  $<$5.59       &                &    \\
165808.33+615001.9 &    0.3742  &    & 18.48 & NOT & 2009-04-04 & 1000 &  0.86$\pm$0.27 &   37.2$\pm$4.1 & S07\\
170108.90+395443.1 &    1.8900? &    & 19.21 &  CA & 2009-02-26 &  720 &  $<$1.20       &                &    \\
170124.64+395437.1 &    0.5071? &    & 16.88 & NOT & 2009-04-04 &  150 &  6.75$\pm$0.21 &    1.0$\pm$1.4 & S07\\
171445.55+303628.0 &    0.8500? &    & 19.11 & NOT & 2009-04-04 &  900 &  $<$0.54       &                & S07\\
171501.36+292912.3 &            &    & 19.49 &  CA & 2009-02-26 &  720 &  $<$4.33       &                &    \\
172640.50+595550.2 &    0.3471? &    & 20.06 &  CA & 2009-02-26 & 1000 &  2.82$\pm$1.12 &  166.6$\pm$4.6 &    \\
173719.12+570216.5 &            &    & 19.89 &  CA & 2009-02-26 &  720 &  6.85$\pm$2.06 &   48.3$\pm$4.2 &    \\
205523.36-050619.3 &    0.3426  &    & 19.00 & NTT & 2008-10-05 &   90 &  3.28$\pm$0.96 &    6.0$\pm$3.9 &    \\
205938.57-003756.0 &    0.3354  &    & 19.29 & NTT & 2008-10-06 &  180 &  $<$3.46       &                &    \\
211552.88+000115.5 &            &    & 19.47 & NTT & 2008-10-05 &  360 &  $<$2.22       &                & P10b\\
211611.89-062830.4 &    0.2916  &    & 18.85 & NTT & 2008-10-04 &   90 &  3.56$\pm$0.85 &    0.6$\pm$3.6 & S07\\
212019.13-075638.4 &            &    & 19.87 & NTT & 2008-10-06 &  360 &  $<$3.16       &                & P10b\\
213950.32+104749.6 &    0.2960  &    & 20.11 & NTT & 2008-10-05 &  600 &  $<$2.37       &                & P10b\\
215051.73+111916.5 &    0.4946? &    & 19.05 & NTT & 2008-10-04 &  180 &  4.88$\pm$0.89 &  157.9$\pm$3.2 &    \\
215305.36-004230.7 &    0.3416  &  Y & 18.26 & NTT & 2008-10-06 &   60 &  5.86$\pm$0.75 &   95.5$\pm$2.7 &    \\
215650.34-085535.4 & $>$1.0179? &  Y & 18.77 & NTT & 2008-10-06 &   90 &  3.24$\pm$1.21 &   74.0$\pm$4.4 &    \\
221108.34-000302.5 &    0.3619  &  Y & 18.80 & NTT & 2008-10-06 &   90 &  6.03$\pm$0.70 &   43.3$\pm$2.6 &    \\
221109.88-002327.5 &    0.4476  &    & 19.64 & NTT & 2008-10-06 &  360 &  5.03$\pm$0.73 &   47.1$\pm$2.9 &    \\
221456.37+002000.1 &            &    & 19.29 & NTT & 2008-10-04 &  180 & 10.70$\pm$1.18 &  156.1$\pm$2.5 &    \\
224448.11-000619.3 &            &    & 19.11 & NTT & 2008-10-05 &  180 &  4.24$\pm$0.60 &  146.7$\pm$2.8 &    \\
224730.19+000006.5 &            &  Y & 19.01 & NTT & 2008-10-05 &   90 &  4.24$\pm$0.52 &  124.1$\pm$2.6 &    \\
224819.44-003641.6 &    0.2123  &    & 18.77 & NTT & 2008-10-04 &   90 &  1.27$\pm$0.57 &  145.8$\pm$4.7 & S07\\
225624.27+130541.7 &            &    & 18.57 & NTT & 2008-10-04 &   90 & 15.56$\pm$0.80 &   19.3$\pm$1.7 & S07\\
231000.81-000516.3 & $>$1.6800? &    & 19.00 & NTT & 2008-10-05 &   90 &  $<$3.21       &                & P10b\\
233445.56+154711.1 &            &    & 19.56 & NTT & 2008-10-04 &  360 & 11.73$\pm$0.82 &   81.9$\pm$2.0 &    \\
235604.03-002353.8 &    0.2830  &  Y & 18.70 & NTT & 2008-10-06 &   90 &  $<$3.48       &                &    \\
\end{longtable}
\tablefoottext{a}{Target has an entry in the NED according to C05.}
\tablefoottext{b}{Object r-band magnitude from the SDSS.}
\tablefoottext{c}{The telescope used for the polarimetry. NTT: New Technology Telescope, La Silla, Chile; CA: Calar Alto 2.2 m telescope, Spain; NOT: Nordic Optical Telescope, La Palma, Spain.}
\tablefoottext{d}{Previous measurements of targets. S07 contains polarimetric observations, P10b a combination of variability, radio and X-ray information.}
\tablebib{(S07)~\citet{2007ApJ...663..118S}, (P10b) \citet{2010ApJ...721..562P}}
}
\end{document}